\documentclass[12pt]{book}

\usepackage[dvips]{graphicx,color}
\usepackage{makeidx,observe,cosmology,epsfig}

\makeauthorindex
\makeindex

\BookTitle{New Trends in Theoretical and Observational Cosmology}
\CopyRight{\copyright 2001 by Universal Academy Press, Inc.}

\begin{document}

\BookTitle{\itshape New Trends in Theoretical and Observational Cosmology}
\CopyRight{\copyright 2001 by Universal Academy Press, Inc.}
\pagenumbering{arabic}

\chapter
{Gravitational Lensing of the Most Distant Known Supernova, SN1997ff}

\author{Christofer GUNNARSSON\\
{\it Fysikum, Stockholm University, SCFAB, SE-106 91 Stockholm, Sweden}}
\AuthorContents{C.\ Gunnarsson}
\AuthorIndex{Gunnarsson}{C.}

\section*{Abstract}
We investigate the effect of gravitational lensing on the farthest
known supernova, 
SN1997ff. The SN was found at $z\sim 1.7$ in
the Hubble Deep Field North and is most likely of Type Ia. Due to our
poor knowledge of the
properties of the lensing foreground galaxies, we conclude that large
magnification effects are possible for reasonable lens parameter values
implying that this single SN does not put any strong constraints on the
cosmological parameters, grey dust obscuration or luminosity evolution
of SNIa until we can model the lensing with high accuracy. 

\section{Introduction}
A primary goal for cosmology is to determine the total energy density of the
Universe, $\Omega$ and its constituents. It has been long recognised that this 
goal is achievable through the study of the redshift-distance relation
of Type Ia supernovae (SNe). The relation is sensitive to different
values of the cosmological parameters and this difference is more
prominent at high redshifts. Therefore, the recently discovered SN at
$z\sim 1.7$ 
\cite{riess2001cgunn} 
might prove to be invaluable in this respect. However, different
systematic effects such as obscuration by grey dust,  
luminosity evolution of Type Ia SNe and gravitational lensing are also
possibly more  
severe at high redshifts. In Ref.~\cite{pointcgunn,rahmancgunn}, 
the systematic effects of gravitational lensing 
on a large sample of SNe have been investigated.
Here, (with more details in Ref.~\cite{uscgunn}), we investigate the
effects of  
gravitational lensing due to galaxies lying close to the line-of-sight to
SN1997ff, generalising the work in Refs.~\cite{lewiscgunn} and
\cite{riess2001cgunn}  
by investigating the combined
effects from a larger number of galaxies and estimating the masses and 
velocity dispersions of the lensing galaxies from the measured luminosities.
Riess et al.~\cite{riess2001cgunn} argued that the observed brightness
of SN1997ff  
suggests that there cannot be a sizeable luminosity evolution for Type Ia's
nor significant extinction by dust. Our work shows that the possible lensing 
magnification effects are large enough that the data is also consistent with 
an intrinsically dimmer supernova, or with significant dust density 
along the line-of-sight. 
 
\section{Method}\label{sec:method}
From the Hubble Deep Field North (HDF-N; \cite{url:hdfncgunn}), 
we obtain the relative positions and redshifts of galaxies lying 
in the proximity of the line-of-sight to SN1997ff. 
All (11) galaxies within $0\leq z \leq 1.7$ and which are
lying closer than 
10 arc-seconds to the line-of-sight
to SN1997ff and its host galaxy have been included in the study. 
We model the matter distribution of the galaxies as truncated isothermal
spheres characterised by their velocity dispersion, $v$, which can be
estimated from the Faber-Jackson
relation
\begin{equation}
  \label{FJ}
  \frac{v}{v_*}=\left(\frac{L}{L_*}\right)^{0.25}=10^{0.1(M-M_*)}.
\end{equation}
Here, $M$ is the absolute magnitude of the galaxy as measured in the $b_J$ 
magnitude system and $L$ is the luminosity. The star indicates typical galaxy values on $v$, $L$
and $M$ (and the mass, $m$, below).
We calculate $M$ by performing cross-filter K-corrections (assuming early-type galaxy spectra) 
on the observed magnitudes in
the $r$ and $i$ bands (ST magnitude system) into rest-frame $b_J$.

To estimate the masses of the lensing galaxies, we combine the observed
luminosities with the mass-to-luminosity ratio \cite{book:Peeblescgunn}
\begin{equation}
  \label{mtol}
  \frac{m}{m_*}=\left(\frac{L}{L_*}\right)^{1.25}=10^{0.5(M_*-M)},
\end{equation}
valid for early-type galaxies lying in the fundamental plane. 
From the velocity dispersion and mass of the galaxies, we can compute
the truncation radii of the halos.
The magnification and deflection of the light from SN1997ff are calculated
by using the multiple lens-plane method
\cite[Ch.~9]{schneidercgunn} where the mass of each of the
lensing galaxies is projected onto a plane at the galaxy
redshift. 
Next, the deflection angle and magnification in each plane is computed
and by tracing the light-rays backwards from the observer to the source,
we can determine the magnification and position of the ray in the
source plane. 
Our calculations have been made using $\Omega_{\rm M} =0.3$, $\Omega_{\Lambda} =0.7$, $h=0.7$ 
and the filled beam approximation when 
calculating cosmological distances.
A typical value of $v_*=238$ km/s was obtained in Ref.~\cite{wilsoncgunn} for
this cosmology.   

\begin{figure}[t]
  \centerline{\epsfig{figure=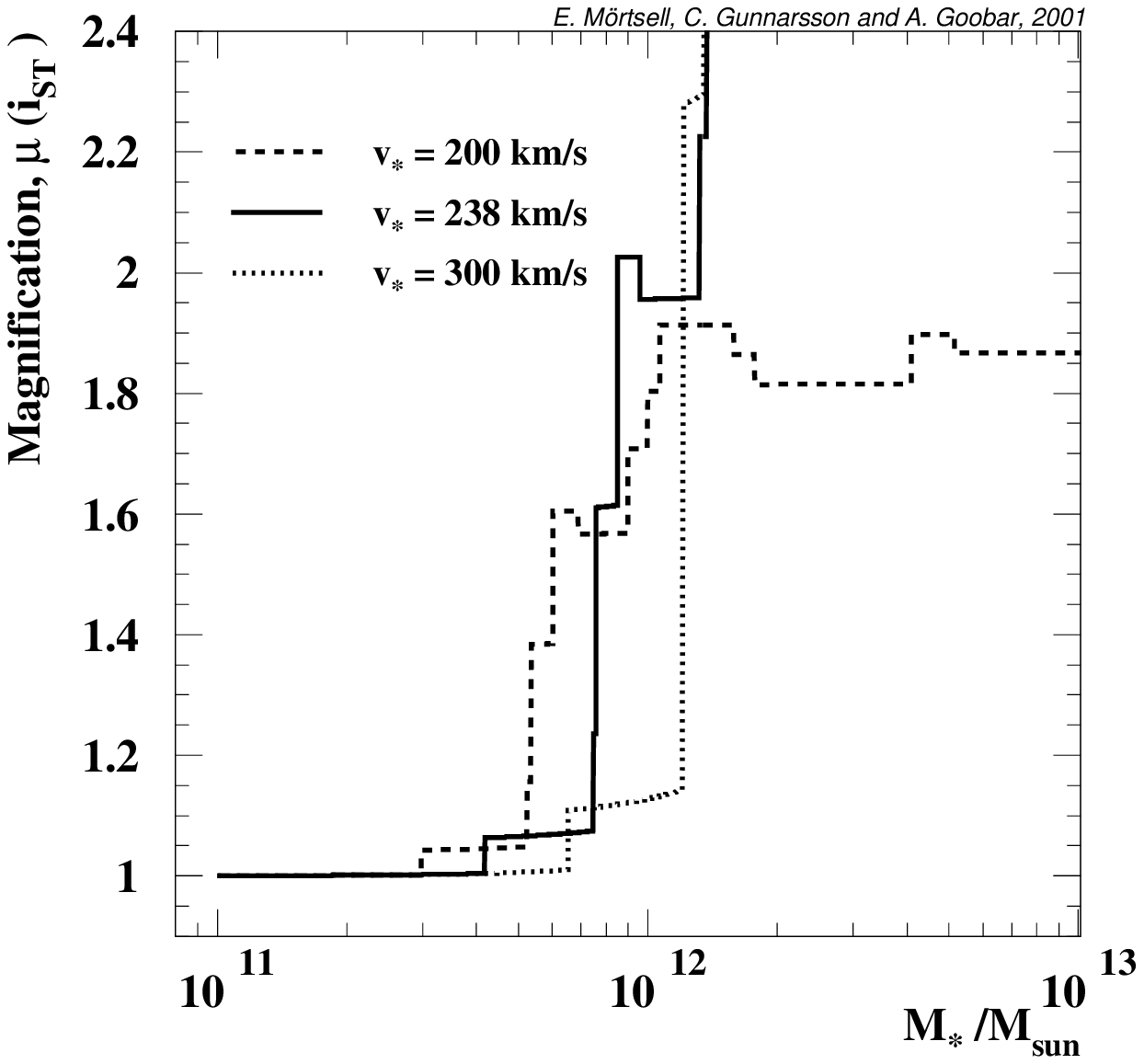,width=0.4\textwidth}
  \epsfig{figure=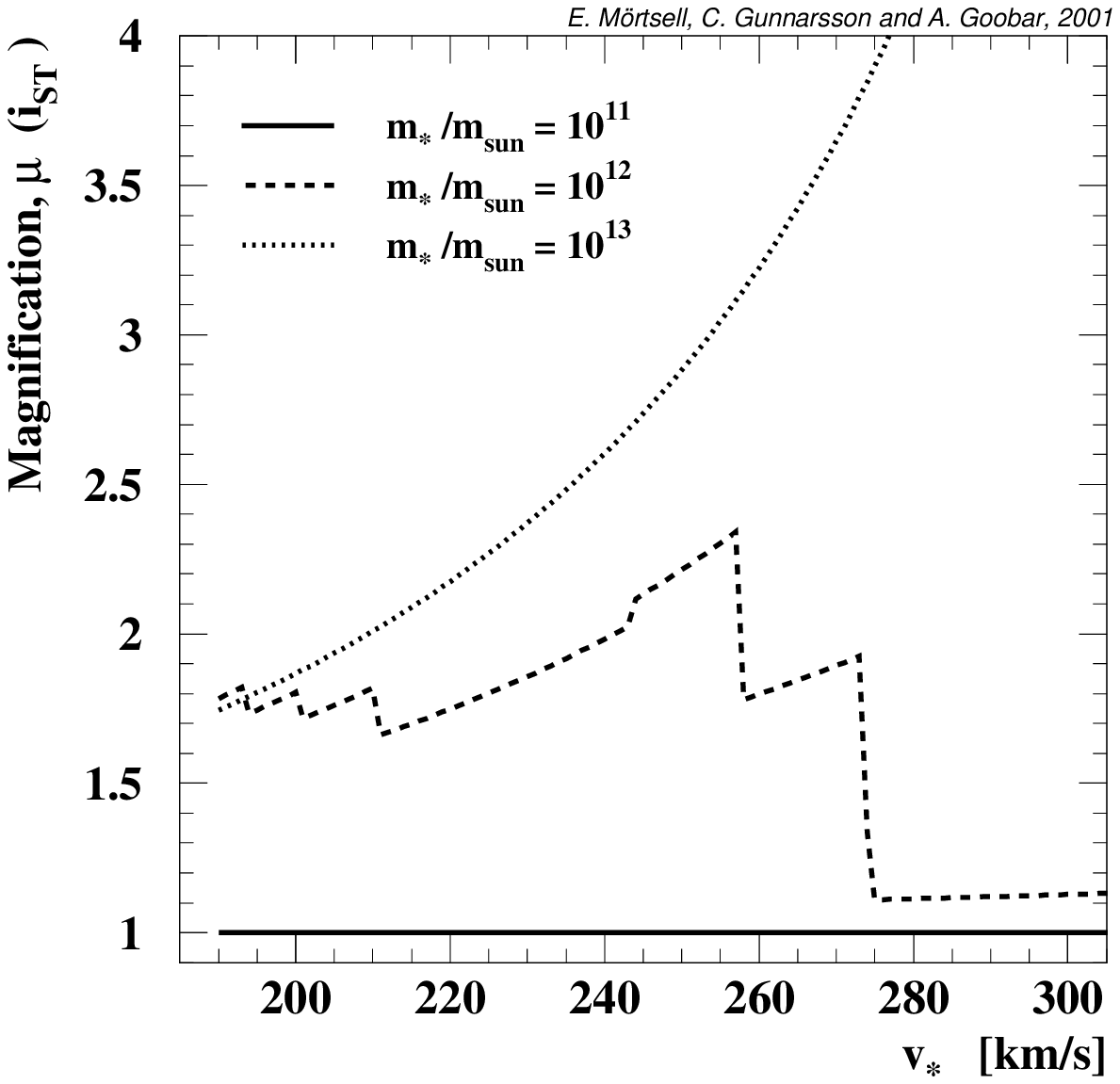,width=0.4\textwidth}}
  \caption{The magnification, $\mu$, as a function of the typical
  galaxy mass (left panel) and $\mu$ as a function of typical galaxy
  velocity dispersion (right panel). Both as obtained from the $i_{\rm
  ST}$ band.}
\label{fig:muvari} 
\end{figure}

\section{Results and discussion}
\label{results}
\subsection{Magnification}
\label{mag}
In Fig.~\ref{fig:muvari}, the magnification of SN1997ff is given as
a function 
of the mass normalisation in the left panel and of the velocity dispersion
normalisation in the right panel. The masses and velocity dispersions
are calculated from the observed luminosities $i_{\rm ST}$-band but
the $r_{\rm ST}$-band give very similar results.   
For a circularly symmetric lens, the magnification is determined
by the mass within  
the radius defined by the impact
parameter of the light-ray and the surface mass density (or
\emph{convergence}) at this radius.  
The effect of the convergence here is to increase the magnification.
Therefore, a light-ray that is passing a plane outside a halo for some
specific values of $m_*$ and $v_*$ might pass inside the same halo
for some other values, thereby gaining the convergence component of
the magnification or vice versa. This is what causes the
discontinuities in the magnification in the plots.   
Of course, these effects are unphysical in the sense that they are very 
sensitive to the specific modeling of the halos, in this case the steepness of
the density profile and the cut-off radii. 
It is also an indication that lensing effects are very model dependent, 
and thus very detailed, individual modeling of the lensing galaxies is
necessary to make robust predictions of the magnification.

\subsection{Clues from host galaxy}\label{sec:clues}
The appearance of the host galaxy might offer some clues to the
magnitude of the  
lensing effect. Riess et al.~\cite{riess2001cgunn} studied the lensing
effects of the foreground galaxy closest to the line-of-sight and
concluded that the probability of a large magnification is small with
such a small ellipticity as is observed (axis ratio $\sim0.85$). For a
single lens, this is true but when including all foreground galaxies
close to the line-of-sight, the situation proves to be much more
complicated. The results in Ref.~\cite{uscgunn} shows that there is
no simple relation 
between magnification and ellipticity when more than one lens is
included in the study.

\section{Conclusions}\label{sec:conclusions}
Our lensing analysis show
that a large range  
of magnifications of SN1997ff is possible for reasonable values of the
galaxy masses and  
velocity dispersions. The value of the magnification is
very sensitive to details in the modelling of the matter distribution in the
lensing galaxies. The apparent (lack of)
ellipticity of the host galaxy can be shown not to put any strong
constraints on the 
magnitude of the magnification effect. 
Thus we conclude, that in order to use the apparent magnitude of a
single high redshift  
SN to infer the values of any cosmological quantities, or even to
place meaningful limits  
on the possible dimming of Type Ia SNe by intergalactic grey dust or
luminosity evolution, very careful modelling of the galaxies  
along the line-of-sight is needed in order to 
control the systematic effects from lensing. 

This work has been carried out together with Edvard M\"ortsell and
Ariel Goobar.
The author would like to thank Lars Bergstr\"om, Joakim Edsj\"o and
Peter Nugent for helpful discussions and
Tomas Dahl\'en for providing galaxy spectral templates. 
I would also like to thank the Swedish Research Council for financial support.







\end{document}